\documentclass{ws-ijmpe}

\begin{document}

\markboth{R. Gonz\'{a}lez Felipe}{Neutrinos and the matter-antimatter asymmetry in the Universe}

\catchline{}{}{}{}{}

\title{Neutrinos and the matter-antimatter asymmetry in the Universe}

\author{\footnotesize R. GONZ\'{A}LEZ FELIPE}

\address{Instituto Superior de Engenharia de Lisboa\\
Rua Conselheiro Em\'{\i}dio Navarro, 1959-007 Lisboa, Portugal\\
Centro de F\'{\i}sica Te\'{o}rica de Part\'{\i}culas, Instituto Superior T\'{e}cnico,\\ Universidade T\'{e}cnica de Lisboa\\
Avenida Rovisco Pais, 1049-001 Lisboa, Portugal\\
gonzalez@cftp.ist.utl.pt}

\maketitle


\begin{abstract}
The discovery of neutrino oscillations provides a solid evidence for nonzero neutrino masses and leptonic mixing. The fact that neutrino masses are so tiny constitutes a puzzling problem in particle physics. From the theoretical viewpoint, the smallness of neutrino masses can be elegantly explained through the seesaw mechanism. Another challenging issue for particle physics and cosmology is the explanation of the matter-antimatter asymmetry observed in Nature. Among the viable mechanisms, leptogenesis is a simple and well-motivated framework. In this talk we briefly review these aspects, making emphasis on the possibility of linking neutrino physics to the cosmological baryon asymmetry originated from leptogenesis.
\end{abstract}

\section{The neutrino puzzle}

Neutrino experiments have provided a convincing evidence that neutrinos oscillate among different flavors. The experimental data collected over more than a decade imply that at least two neutrinos have nonzero mass and that there is mixing in the lepton sector, in analogy to the Cabibbo-Kobayashi-Maskawa (CKM) mixing in the quark sector. Yet neutrinos have surprised us since their properties are quite different from those of charged fermions. Table 1 reports our present knowledge of neutrino mass and mixing parameters obtained by the analysis of the global $3\nu$ oscillation data.~\cite{GonzalezGarcia:2010er,Schwetz:2011qt,Fogli:2011qn} Although neutrino oscillation experiments are not sensitive to the absolute neutrino mass scale, direct kinematical searches and cosmological bounds set $m_\nu \lesssim \mathcal{O}(1)$~eV. Thus neutrinos are at least six orders of magnitude lighter than the other fermions in the Standard Model (SM). Their mass hierarchies among generations also turn out to be different from those of charged leptons and quarks. In particular, there is a mild neutrino hierarchy $m_3/m_2 < 6$. By contrast, for the charged-lepton masses, $m_\tau/m_\mu \sim 17$ and $m_\mu/m_e \sim 207$, and the quark mass relations $m_t/m_c \sim 135, m_c/m_u \sim 510, m_b/m_s \sim 40, m_s/m_d \sim 20$ are verified.

\begin{table}[pt]
\tbl{Best-fit values with $1\sigma$ errors for the three-flavor neutrino oscillation parameters determined by three different analyses of the global neutrino oscillation data. Here $\Delta m^2_{ij} \equiv m_i^2-m_j^2$ are the neutrino mass-squared differences ($m_i$ is the mass of the eigenstate $\nu_i$), and $\theta_{ij}$ are the mixing angles. When available, the upper (lower) row refers to a normal (inverted) neutrino mass hierarchy $m_1<m_2<m_3$ ($m_3<m_1<m_2$).}
{\begin{tabular}{lccc} \toprule
 & Gonzalez-Garcia et al.~\cite{GonzalezGarcia:2010er} &Schwetz et al.~\cite{Schwetz:2011qt}  &Fogli et al.~\cite{Fogli:2011qn}
\\ \colrule
$\Delta m^2_{21}\,[10^{-5}$ eV$^{2}$] &$7.59\pm 0.20$
&$7.59^{+0.20}_{-0.18}$ & $7.58^{+0.22}_{-0.26}$
\\ \\
$\Delta m^2_{31}\,[10^{-3}$ eV$^{2}$] &$2.46\pm 0.12$
&$2.50^{+0.09}_{-0.16}$ & $2.35^{+0.12}_{-0.21}$\\
 & $-2.36\pm 0.11$\hphantom{0} & $-2.40^{+0.09}_{-0.08}$\hphantom{0}
& $-2.35^{+0.21}_{-0.12}$\hphantom{0}\\ \\
$\sin^2\theta_{12}$ &\hphantom{00}$0.319\pm 0.016$
&\hphantom{00}$0.312 ^{+0.017}_{-0.015}$ &\hphantom{00}$0.312^{+0.017}_{-0.016}$\\  \\
$\sin^2\theta_{23}$ & $0.46^{+0.08}_{-0.05}$\hphantom{0}
&$0.52^{+0.06}_{-0.07}$ &$0.42^{+0.08}_{-0.03}$\\
 & &\hphantom{00}$0.52\pm 0.06$
&\\ \\
$\sin^2\theta_{13}$ & \hphantom{0} $0.0095^{+0.013}_{-0.007}$
&\hphantom{00} $0.013^{+0.007}_{-0.005}$ &\hphantom{000}$0.025\pm 0.007$\\
 & &\hphantom{00} $0.016^{+0.008}_{-0.006}$ &\\
\botrule
\end{tabular}}
\end{table}

The mixing pattern in the lepton sector also looks quite distinct from its analogue in the quark sector. Indeed, while the CKM matrix exhibits a small mixing pattern, the Pontecorvo-Maki-Nakagawa-Sakata (PMNS) lepton mixing matrix has two large mixing angles: the solar $\theta_{12} \simeq 34^\circ$ and atmospheric $\theta_{23} \simeq 45^\circ$. The third angle $\theta_{13}$ is mainly constrained by the reactor data to be small, $\theta_{13} \lesssim 12^\circ$ and, from the combined global data, there is a hint for a nonzero $\theta_{13}$. Recent data from T2K~\cite{Abe:2011sj} and MINOS~\cite{Adamson:2011qu} experiments also indicate a relatively large value for $\theta_{13}$. At 90\% C.L., the T2K data are consistent with $0.03\, (0.04)< \sin^2 2\theta_{13} < 0.28\, (0.34)$ for normal (inverted) hierarchy in the absence of Dirac CP violation. The MINOS collaboration reports the best-fit values $2\sin^2(\theta_{23})\sin^2(2\theta_{13})\,\mathord{=}\,0.041^{+0.047}_{-0.031}\ \bigl(0.079^{+0.071}_{-0.053}\bigr)$. If these results are confirmed, they will have strong impact on neutrino physics and, particularly, will open the window for the possibility of detecting leptonic CP violation.

Another puzzling aspect of neutrinos is their nature. In the SM neutrinos are strictly massless. This is so because no right-handed (RH) neutrino fields are introduced, thus preventing a Dirac mass term for them. No Majorana mass term of the form $m \nu^T_L C^{-1} \nu_L$ can be either generated since such term would violate $B-L$, which is an exact symmetry in the SM, conserved at the quantum level. If RH neutrino fields are added to the SM particle content, neutrinos could then be either Dirac or Majorana particles. Yet, theoretically, the latter appear to be more natural. Indeed, a Dirac neutrino mass would require extremely small Yukawa couplings $y_\nu \lesssim 10^{-12}$. On the other hand, a Majorana mass term can be interpreted as the lowest-order effective operator beyond the SM, namely, the dimension-five Weinberg operator $y_\nu^2\, (\ell \phi) (\ell \phi)/M$, where $\phi$ is the SM Higgs doublet, $\ell$ stands for the lepton doublet, and $M$ is a mass scale associated to some new physics.~\cite{Weinberg:1980bf} Then, after electroweak symmetry breaking, neutrinos would acquire a Majorana mass $m_\nu \sim y_\nu^2 v^2/M$, which can be easily suppressed if the scale $M$ is much higher than the electroweak scale $v$, without the need of unnaturally small Yukawa couplings. This approach is commonly referred to as the seesaw mechanism for neutrino masses.

\subsection{The seesaw solution}

The seesaw is a simple and attractive mechanism for the realization of the Weinberg operator in gauge theories. In this framework, the effective operator is induced by the exchange of mediator particles with a mass scale $M$. Depending on the SU(2) representation of the mediators, several seesaw realizations are conceivable for neutrino mass generation. The most common are the type-I (singlet RH neutrinos), type-II (triplet scalars) and type-III (triplet fermions) seesaw mechanisms. Such representations are commonly present in grand unified theories. For instance, RH neutrinos nicely fit into the \textbf{16} spinorial representation of SO(10).

Introducing in the SM right-handed neutrinos $\nu_{Ri}$ with heavy masses $M_i$ is one of the simplest possibilities to generate light neutrino masses. To be specific, let us consider the case of three RH neutrinos, i.e. one for each fermion generation.~\footnote{Current neutrino data are also consistent with the existence of just two right-handed neutrinos.} The Lagrangian reads
\begin{align}\label{LtypeI}
    \mathcal{L}=\mathcal{L}_\text{SM} + i\,\overline{\nu}_{Ri} \gamma_\mu \partial^\mu \nu_{Ri}  -\mathbf{Y}_{\alpha i}^{\nu \ast}\, \bar{\ell}_{\alpha} \tilde{\phi}\, \nu_{Ri} - \frac{1}{2}\, \overline{\nu}_{Ri}\, (\mathbf{m}_R)_{ij}\, \nu_{Rj}^c + \text{H.c.},
\end{align}
where $i=1,2,3,\, \alpha=e, \mu, \tau$; $\tilde{\phi} = i \sigma_2 \phi^\ast$, $\mathbf{Y}^\nu$ is the $3\times 3$ Dirac-neutrino Yukawa coupling matrix and $\mathbf{m}_R$ is the $3\times 3$ symmetric RH neutrino mass matrix. Integrating out the heavy Majorana fields in Eq.~\eqref{LtypeI}, and after the electroweak symmetry breaking, the effective neutrino mass matrix is given by the seesaw relation
\begin{align}\label{mnutypeI}
    \mathbf{m}_\nu = - v^2\, \mathbf{Y}^\nu\, \mathbf{m}_R^{-1}\, \mathbf{Y}^{\nu T}.
\end{align}
Therefore, for sufficiently large Majorana masses $M_i$, the required light neutrino masses $m_i$ can be naturally reproduced. In the basis where the charged-lepton Yukawa couplings are diagonal, the matrix $\mathbf{m}_\nu$ is diagonalized by the $3\times 3$ unitary PMNS leptonic mixing matrix $\mathbf{U}$,
\begin{align}\label{dnutypeI}
    \mathbf{U}^T\, \mathbf{m}_\nu\, \mathbf{U} = \text{diag}(m_1,m_2,m_3).
\end{align}

Light neutrino masses can be generated from other high-scale seesaw mechanisms in a similar manner. As a bonus, seesaw models also give the possibility of implementing leptogenesis,~\cite{Fukugita:1986hr} i.e. the generation of a lepton asymmetry in early Universe due to the out-of-equilibrium and CP-violating decays of the seesaw mediators. This asymmetry will be subsequently converted into a baryon asymmetry, thus offering a dynamical explanation for the absence of any primordial antimatter in the observable Universe.

\subsection{The flavor solution}

The neutrino oscillation data in Table~1 hint at certain symmetry properties in the leptonic mixing. A remarkable fact is that the $\nu_2$ mass eigenstate is quite equally mixed with all three flavors $\nu_{e,\mu,\tau}$ (trimaximal mixing), whereas $\nu_3$ is almost equally mixed with $\nu_\mu$ and $\nu_\tau$ (bimaximal mixing) and contains a very small $\nu_e$ component. This peculiar flavor structure has led many authors to look for underlying flavor symmetries compatible with the data.~\cite{Altarelli:2010gt} Discrete groups such as $A_4$, $S_3$, $S_4$, $D_4$, $D_7$, $A_5$, $T^\prime$, $T_7$, and $\Delta(27)$ have been widely used in several proposals.~\cite{Smirnov:2011jv} In this context, the permutation group $A_4$ has been quite popular since it is particularly suitable to reproduce the tribimaximal mixing~\cite{Harrison:2002er}
\begin{align} \label{UTBM}
\mathbf{U}_\text{TB}=\begin{pmatrix}
\hphantom{0}\sqrt{\frac{2}{3}}&\hphantom{00}\sqrt{\frac{1}{3}}&\hphantom{000}0\\
-\sqrt{\frac{1}{6}}&\hphantom{00}\sqrt{\frac{1}{3}}&\hphantom{0}-\sqrt{\frac{1}{2}}\\
-\sqrt{\frac{1}{6}}&\hphantom{00}\sqrt{\frac{1}{3}}&\hphantom{000}\sqrt{\frac{1}{2}}
\end{pmatrix},
\end{align}
which is consistent at $(1-2)\sigma$ level with the experimental data.

An attractive feature of mass-independent mixing schemes such as $\mathbf{U}_\text{TB}$ is that they lead to a predictive neutrino mass matrix, which contains just a few parameters that can be directly related to low-energy neutrino observables (e.g. the neutrino mass-squared differences, the absolute neutrino mass scale, and the effective mass parameter in neutrinoless double beta decays). It is nevertheless fair to say that constructing neutrino mass models based on flavor symmetries is not an easy task. Additional discrete and/or continuous symmetries are typically required to guarantee the correct vacuum alignment and mass hierarchies, thus making these models quite intricate.

In the light of the recent T2K results, models that lead to tribimaximal mixing appear to be disfavored. In general, deviations from $\theta_{13}=0$ in these models cannot bring $\theta_{13}$ into agreement with data without spoiling the predictions for the solar and atmospheric mixing angles. Several alternative solutions have been recently put forward to explain a relatively large value of $\theta_{13}$ in the framework of discrete flavor symmetries.~\cite{Ma:2011yi,Meloni:2011fx,Toorop:2011jn}

It is also remarkable that the imposition of certain flavor symmetries in the lepton sector of the theory may lead to severe constraints on the leptonic CP asymmetries relevant for leptogenesis in the framework of seesaw models. In particular, in type-I and type-III seesaw flavor models that lead to an exact mass-independent leptonic mixing the leptogenesis CP asymmetries are zero in leading order.~\cite{Jenkins:2008rb,Bertuzzo:2009im,AristizabalSierra:2009ex,Felipe:2009rr} By contrast, in a type-II seesaw framework the leptonic CP asymmetries are in general nonzero and leptogenesis is viable.~\cite{Varzielas:2011tp}

\section{The cosmological puzzle}

No primordial antimatter is found in our observable Universe. From the analysis of the Wilkinson Microwave Anisotropy Probe (WMAP) seven-year data combined with baryon acoustic oscillations it is inferred that~\cite{Komatsu:2010fb}
\begin{align}\label{etaB}
    \eta_B \equiv \frac{n_B-n_{\bar{B}}}{n_\gamma}=(6.20 \pm 0.15) \times 10^{-10},
\end{align}
where $n_B, n_{\bar{B}}$ and $n_\gamma$ are the number densities of baryons, antibaryons and photons, respectively. Remarkably, this result is also consistent with the one obtained from the concordance of the light elements and big bang nucleosynthesis. The explanation of such a tiny but nonzero number poses a challenge to both particle physics and cosmology. WMAP measurements have also made it clear that the current state of the Universe is very close to a critical density and that the primordial density perturbations that seeded large-scale structure formation are nearly scale invariant and Gaussian, which is consistent with the inflationary paradigm. Since any primordial asymmetry would have been exponentially wiped out during the inflationary period, one then expects the baryon asymmetry to be generated by some dynamical mechanism after inflation. The necessary ingredients for such a dynamical recipe have been formulated long ago by Sakharov: B-violation, C and CP-violation, and departure from thermal equilibrium.~\cite{Sakharov:1967dj}

It is noteworthy that the SM contains the three Sakharov ingredients. Yet not all of them are available in a sufficient amount. Baryon number is violated by the electroweak sphaleron processes, which are fast and unsuppressed in early Universe. The C symmetry is maximally violated by the weak interactions, and CP is violated by the CKM phase, but not enough to generate the required asymmetry $\eta_B$. Finally, at the electroweak phase transition departure from thermal equilibrium takes place. However, a successful baryogenesis requires a strongly first order phase transition, which can only occur if the Higgs mass is less than 60~GeV, i.e. in a mass range that is already excluded by the experimental electroweak data. Thus the explanation of the cosmological baryon asymmetry requires new physics beyond the SM.

\begin{figure}[t]
\centerline{\psfig{file=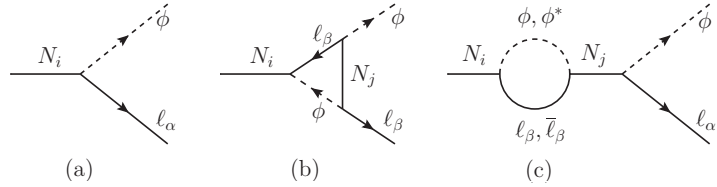,width=10cm}}
\caption{\label{fig1} Feynman diagrams contributing to the CP asymmetries $\epsilon_i^{\alpha}$ in
type-I seesaw leptogenesis: (a) Tree-level graph, (b) one-loop vertex graph, and (c) one-loop self-energy graphs (the diagram with an internal $\ell_\beta$ is lepton flavor and lepton number violating, whereas the one with an internal $\bar{\ell}_\beta$ is lepton flavor violating but lepton number conserving).}
\end{figure}

\subsection{The leptogenesis solution}

Among the viable dynamical scenarios to explain the matter-antimatter asymmetry, leptogenesis is undoubtedly one of the simplest, most attractive and well-motivated mechanisms.~\cite{Fukugita:1986hr,Davidson:2008bu} New heavy particles are introduced in the theory in such a way that the interactions relevant for leptogenesis are simultaneously responsible for the seesaw neutrino masses. The three Sakharov conditions are naturally fulfilled in this framework: (i) the seesaw mechanism requires lepton number violation and nonperturbative $(B+L)$-violating sphaleron processes will partially reprocess the lepton asymmetry into a baryon asymmetry; (ii) neutrino complex Yukawa couplings provide the necessary source of CP violation; (iii) departure from thermal equilibrium is accomplished by the out-of-equilibrium decays of the new heavy particles at temperatures above the electroweak scale.

As an illustrative example, let us consider again the case of the SM extended with three RH neutrinos. The relevant interactions are described by the type-I seesaw Lagrangian of Eq.~\eqref{LtypeI}. Working in the mass eigenbasis of the charged leptons $\ell_\alpha$, and in the mass eigenbasis $N_i$ of the RH neutrinos $\nu_{Ri}$, the CP asymmetry $\epsilon_i^\alpha$ in the lepton flavor $\alpha$ produced in the $N_i$ decays is given by
\begin{align}\label{eflav}
          \epsilon_i^\alpha & \equiv \frac{\Gamma(N_i\rightarrow \phi\, \ell_\alpha)-\Gamma(N_i\rightarrow \phi^\dagger\, \bar{\ell}_\alpha)}{\sum_\beta \left[\Gamma(N_i\rightarrow \phi \, \ell_\beta)+\Gamma(N_i\rightarrow \phi^\dagger\, \bar{\ell}_\beta)\right]}\nonumber \\
         & = \frac{1}{8\pi\mathbf{H}_{ii}^{\nu}}\sum_{j \neq i} \Bigl\{ \text{Im}\left[\mathbf{Y}^{\nu\ast}_{\alpha i} \mathbf{H}^\nu_{ij} \mathbf{Y}^\nu_{\alpha j}\right]\,\bigl(f(x)
         + g(x)\bigr) + \text{Im}\left[\mathbf{Y}^{\nu\ast}_{\alpha i}\mathbf{H}^\nu_{ji} \mathbf{Y}^\nu_{\alpha j}\right] g^\prime(x)\Bigr\},
    \end{align}
where $\mathbf{H}^\nu \equiv \mathbf{Y}^{\nu\dagger}\mathbf{Y}^\nu$, $x \equiv M_j^2/M_i^2$,
\begin{align}\label{loopfunc}
  f(x) =\sqrt{x}\left[1-(1+x)\ln\left(1+x^{-1}\right)\right], \,
  g(x) = \frac{\sqrt{x}\,(1-x)}{(x-1)^2+\Gamma_{N_j}^2/M_i^2}=g^\prime(x)\sqrt{x},
\end{align}
and $\Gamma_{N_i}= \mathbf{H}^\nu_{ii}M_i/(8\pi)$ is the total tree-level decay rate of $N_i$. The CP asymmetry $\epsilon_i^\alpha$ is obtained from the interference of the tree-level and one-loop diagrams of Fig.~\ref{fig1}. Summing over the lepton flavors, one obtains the unflavored asymmetry
\begin{align}\label{eunflav}
       \epsilon_i = \sum_\alpha \epsilon_i^\alpha = \frac{1}{8\pi\mathbf{H}_{ii}^{\nu}}\sum_{j \neq i} \text{Im}\left[(\mathbf{H}_{ij}^{\nu})^2\right] \,\bigl(f(x) + g(x)\bigr).
\end{align}

If the Universe reheats to a thermal bath composed of particles with gauge interactions after inflation, the final baryon-to-photon number ratio $\eta_B$ can be estimated as the product of three suppression factors: (the leptonic CP asymmetry $\epsilon_i$ in $N_i$-decays) $\times$ (an efficiency factor $\kappa_i$ due to washout processes in scattering, decays and inverse decays) $\times$ (a reduction factor due to chemical equilibrium, charge conservation and the redistribution of the asymmetry among different particle species). In particular, the final efficiency factor $\kappa_i$ is computed by solving numerically the relevant Boltzmann equations, which describe the out-of-equilibrium dynamics of the processes responsible for leptogenesis. Simple analytical estimates can be obtained in some specific regimes.~\cite{Giudice:2003jh,Buchmuller:2004nz}

Traditionally, the unflavored regime \eqref{eunflav} has been considered, assuming also a heavy Majorana neutrino mass hierarchy $M_1 \ll M_{2,3}$ (the so-called $N_1$-dominated scenario). In this oversimplified setup, one can show that the CP asymmetry has an upper bound, $|\epsilon_1| \lesssim 3 M_1(m_3 - m_1)/(16\pi v^2)$. To reproduce the observed value given in Eq.~\eqref{etaB}, two constraints must then be satisfied: a lower bound on $M_1$ and the reheating temperature $T_\text{rh}$ independent of the initial conditions, $M_1 \gtrsim  T_\text{rh} \gtrsim 10^9$~GeV, and an upper bound on the light neutrino mass scale, $m \lesssim 0.15$~eV.~\cite{Giudice:2003jh,Buchmuller:2004nz} Another notable feature of the unflavored asymmetry \eqref{eunflav} is that it does not depend directly on the parameters of the PMNS mixing matrix $\mathbf{U}$. This raises the question of whether there is a direct link between leptogenesis and low-energy neutrino observables. As it turns out, any link can only be established in a model-dependent way, and imposing certain conditions on the Yukawa coupling matrix $\mathbf{Y}^\nu$.~\cite{Branco:2001pq,Branco:2002kt,Branco:2002xf}

Going beyond the simple $N_1$-dominated approximation, it is possible to relax the bounds on $M_1$ and $T_\text{rh}$. For instance, if $|M_j - M_i| \simeq 1/2\, \Gamma_{N_j}$, i.e. when the RH neutrinos are quasi-degenerate in mass, the flavored  $\epsilon_i^\alpha$ and unflavored $\epsilon_i$ asymmetries are resonantly enhanced by the self-energy corrections depicted in Fig.\ref{fig1},
\begin{align}\label{eresflav}
           \epsilon_{i}^\alpha
          \simeq -\sum_{j \neq i}  \frac{\text{Re}\,\left[\mathbf{H}^\nu_{ij}\right]\, \text{Im}\left[\mathbf{Y}^{\nu\ast}_{\alpha i}  \mathbf{Y}^\nu_{\alpha j}\right]}{\mathbf{H}^\nu_{ii}\mathbf{H}^\nu_{jj}}, \quad \epsilon_{i} \simeq -\frac{1}{2} \sum_{j \neq i} \frac{\text{Im}\left[(\mathbf{H}_{ij}^{\nu})^2\right]} {\mathbf{H}^\nu_{ii}\mathbf{H}^\nu_{jj}}.
\end{align}
Although theoretically challenging, it is possible to construct models in which the heavy Majorana neutrino mass splitting is naturally as small as the decay width at the leptogenesis scale. In the radiative resonant leptogenesis scenario, the required splitting can be generated by renormalization group running effects, assuming that the heavy Majorana neutrinos are exactly degenerate at the GUT scale.~\cite{GonzalezFelipe:2003fi,Branco:2005ye,Branco:2009by}

Recently, it has also been emphasized that lepton flavor effects can play a significant role in leptogenesis.~\cite{Abada:2006fw,Nardi:2006fx,Abada:2006ea,Pilaftsis:2009pk} When the interactions mediated by the charged lepton Yukawa couplings are in thermal equilibrium, the flavored leptonic asymmetries and the Boltzmann equations for individual flavor asymmetries must be taken into account. Since interactions involving the $\tau$ ($\mu$) Yukawa couplings are in equilibrium for $T \lesssim 10^{12}\,(10^{9})$~GeV, the corresponding lepton doublets are distinguishable mass eigenstates below these temperatures and should be properly introduced into the dynamics. In the flavored regime, the arguments leading to the upper bound on $\epsilon_i$ do not apply, and the leptogenesis efficiency factors can be larger than in the unflavored case. Furthermore, the parameters in the mixing matrix $\mathbf{U}$ directly affect the final baryon asymmetry so that it becomes possible to have successful leptogenesis just from low-energy leptonic CP violation.\cite{Pascoli:2006ie,Branco:2006ce}

\section{The road ahead}

The possibility of linking neutrino physics to the cosmological matter-antimatter asymmetry obviously raises the question about the testability of the seesaw and leptogenesis mechanisms. Since heavy states and/or very small Yukawa couplings are typically present, achieving this goal is not straightforward. There is however a hope that colliders such as the LHC will shed some light on the Majorana nature of neutrinos and, in some cases, probe the seesaw as well. Smoking gun signatures would be the detection of neutrinoless double beta decay and the production of same-sign lepton pairs at colliders.~\cite{Senjanovic:2010nq} Testing leptogenesis seems more difficult, and the success of this task will depend on the energy scale at which leptogenesis takes place and the type of seesaw related to it.~\cite{Frere:2008ct,Blanchet:2009bu} In any case, the upcoming years are expected to be another exciting golden era for neutrino physics, and future experiments may provide important clues to some of the still unanswered questions.

\section*{Acknowledgements}
This work was supported by \textit{Funda\c{c}\~{a}o para a Ci\^{e}ncia e a Tecnologia} (FCT, Portugal) under the projects CERN/FP/116328/2010, PTDC/FIS/098188/2008, and CFTP-FCT Unit 777, which are partially funded through POCTI (FEDER).

\end{document}